\documentstyle[11pt,aaspp4,psfig]{article}

\newcommand{\lsim}{\raisebox{-5pt}{$\;\stackrel{\textstyle<}{\sim}\;$}}

\slugcomment{To appear in Ap.J.}

\lefthead{Vladilo et al.}
\righthead{  
 Metallicity redshift relation 
of Damped Ly\,$\alpha$ systems  }

\begin{document}

\title
{Zinc as a tracer of metallicity evolution of Damped Ly\,$\alpha$ systems }

\author{ Giovanni Vladilo \altaffilmark{1},
Piercarlo Bonifacio \altaffilmark{1}, Miriam Centuri\'on \altaffilmark{1}, and
Paolo Molaro \altaffilmark{1}  }
\affil{$^1$ Osservatorio Astronomico di Trieste, Via G.B. Tiepolo 11, 34131 Trieste, Italy}


\begin{abstract}
Zinc is  a good indicator   of metallicity in Damped Lyman $\alpha$ (DLA) systems
because it is almost unaffected by dust depletion. However, the use of zinc
as a tracer of metallicity evolution has been hampered by the difficulty of
detecting the Zn II resonance lines at high redshift. 
The   measurement of zinc abundance in a DLA system 
at $z_{\rm abs} > 3$ 
obtained by means of the UVES spectrograph at the VLT 
prompted us to re-analyse the full sample of zinc abundances
present in the literature to search for a metallicity-redshift relation in DLA systems.
The study of the  metallicities  of individual systems  
shows   evidence for  an anti-correlation between [Zn/H] and redshift
supported by different types of statistical tools. 
The zinc metallicity decreases by 
$  -0.3  \pm 0.1 $ dex per unit redshift interval in the range 
 $0.5 \lsim z_{\rm abs} \lsim 3.5$.  
This rate  is in good agreement with that found  by Savaglio, Panagia \& Stiavelli (2000)
in their recent study of  DLA abundances corrected   for dust depletion.
The present result does not require   a knowledge of   the dust depletion pattern(s)
in DLA systems. 
On the other hand,
the analysis of the column-density weighted metallicity of the sample, $<Z>$,
does not show  a clear evidence for redshift evolution, 
consistent with previous studies of zinc abundances. 
We propose that the apparent lack of evolution of  $<Z>$ is due  to 
the combination of  selection bias effects together with
the extreme sensitivity of $<Z>$ to low-number statistics.  
\end{abstract}


\section{Introduction}

The redshift-metallicity relation of QSO absorption systems
is a fundamental probe of the chemical evolution of  the universe. 
The study of such relation in Damped Lyman $\alpha$ (DLA) 
systems\footnote{The  QSO absorption systems with neutral hydrogen column density
$N$(HI) $> 2 \times 10^{20}$ atoms cm$^{-2}$ are called Damped 
Lyman $\alpha$ systems owing to the presence of radiative damping wings
in their Lyman $\alpha$ absorption profiles.}
probes, in particular,  the metal enrichment of  the associated
galaxies  located at cosmological distances along the QSO line of sight
(Lu et al. 1996). 
Abundance studies of DLA systems can be used to trace the chemical 
evolution of galaxies in the early universe, starting from the redshifts of the most distant
QSOs.  

 Abundance determinations for different elements are presently available for about 60  
 DLA systems, but the study of their redshift evolution is hampered    
by two main difficulties. One is the uncertainty  of the  abundance  
measurements owing to the possible effects of dust depletion. 
The other is the limited redshift coverage of the sample, which is not always adequate
for probing the presence of evolution. 
One approach to tackle the first difficulty is to   correct the observed abundances
for dust depletion effects (Vladilo 1998, Savaglio et al. 2000). Another approach 
is to use an element unaffected by dust
depletion as a tracer of metallicity evolution (Pettini et al. 1997, 1999). 
Zinc is known to have little affinity with dust since it is  essentially undepleted 
in the interstellar medium (Roth \& Blades 1995). 
Abundance measurements in  Galactic metal-poor stars yield 
[Zn/Fe]\footnote{We adopt the usual convention
[X/Y] = log $N$(X)/$N$(Y) $-$ log (X/Y)$_{\sun}$}
$\approx 0$ 
(Sneden, Gratton \& Crocker 1991), suggesting that zinc is a good tracer of iron. 
The quality of available stellar data leaves open the possibility 
that  new, more precise measurements may reveal
small deviations of [Zn/Fe] from the solar ratio such as those found for the iron-peak elements
Cr, Mn and Co (Ryan, Norris \& Beers 1996).
However,  even in this case, zinc would still be  
a good  indicator of metallicity, even if not a perfect tracer of Fe.

Studies of zinc abundances in DLA systems have not revealed  
evidence for   evolution of the column-density weighted 
 [Zn/H] metallicity  (Pettini et al. 1997, 1999). 
One difficulty in detecting evolution
is  the lack of measurements   at
$z_{\rm abs} > 3$, when the redshifted Zn\,II resonant doublet
$\lambda_{\rm rest} 2025, 2062$ \AA\ falls 
in the reddest part of the visible spectrum.   
In addition, the column-density weighted metallicity is more
prone to be affected by low number statistics than  the unweighted metallicity, as we discuss
in Section 3. 
Here we present the results of a search for redshift evolution performed
by considering  both the
unweighted and the weighted metallicities (Sections 2 and 3, respectively)
and based on the sample of 
[Zn/H]  literature data,   which includes now our recent
measurement at $z_{\rm abs} \simeq 3.4$ (Molaro et al. 2000).

\section{ The anti-correlation between [Zn/H] and redshift }
 
\begin{figure}
\psfig{figure=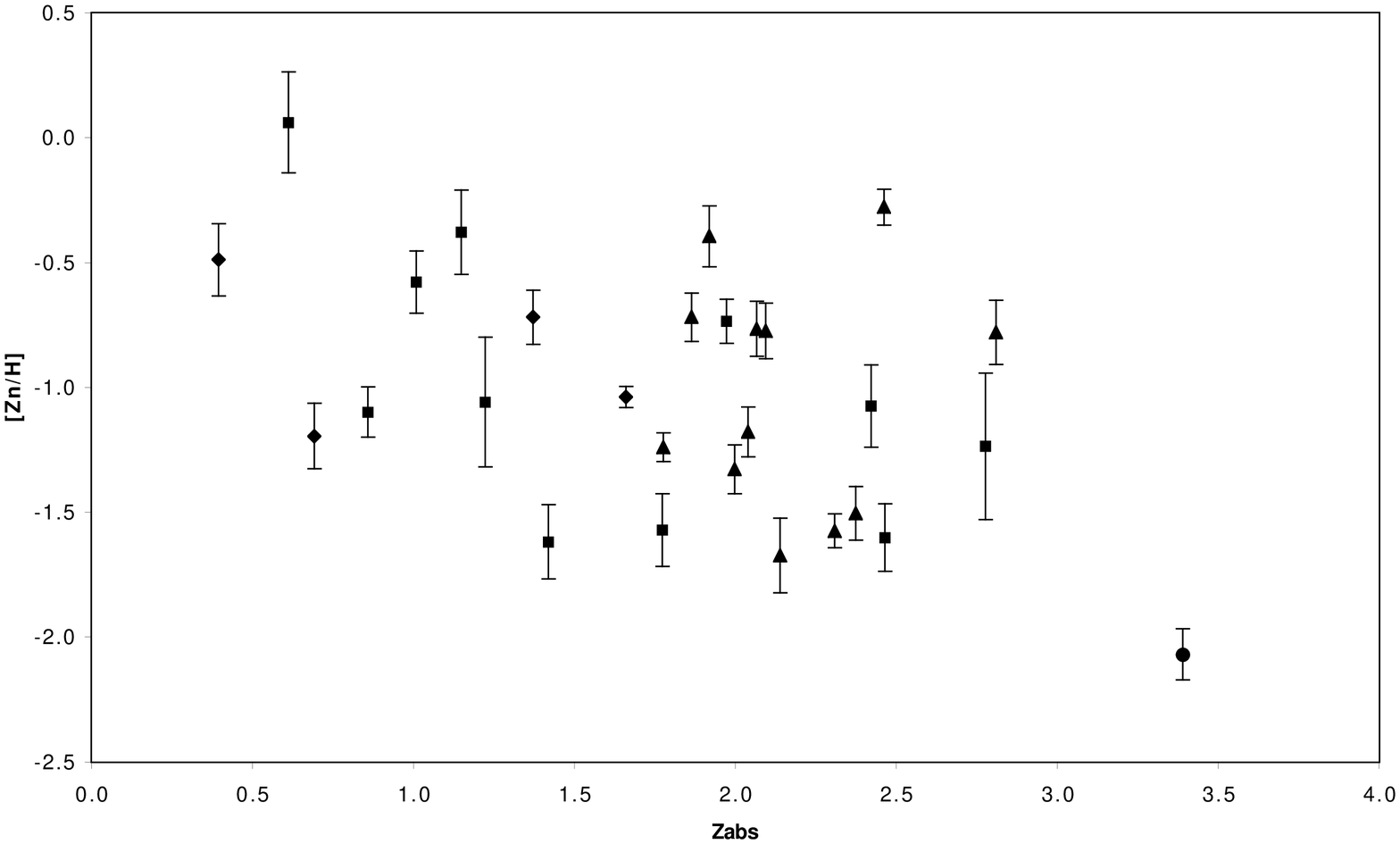,width=17.1cm}
\caption{ Zinc abundances versus absorption redshift in DLA systems.
Squares: data taken from Pettini et al. (1994,1997,1999,2000). 
Triangles: data from Lu et al. (1996) and from
Prochaska \& Wolfe (1996, 1997, 1999). 
Diamonds: data from Meyer \& York (1992), Meyer, Lanzetta \& Wolfe (1995),
 Boiss\'e et al. (1998), and L\'opez et al. (1999).     
Circle:  $z_{\rm abs}=3.3901$ system in QSO 0000-26 (Molaro et al. 2000). }
\label{Fig1}
\end{figure}

In Fig. \ref{Fig1} we show  the metallicity
  [Zn/H] 
in DLA systems
derived from  the Zn\,II and H\,I column density determinations  
currently available in literature. 
References to the original works are indicated in the figure caption. 
The  UVES measurement of the $z_{\rm abs}=3.3901$ system in QSO 0000-26
is the big circle at the right bottom of the figure.
When necessary, the zinc column densities were redetermined
by using the  oscillator strengths by  Bergeson \& Lawler (1993)
which are adopted   for the full sample. 
The   meteoritic abundance by number  $\log ({\rm Zn/H})_{\odot} = -7.33 \pm 0.04$
(Grevesse, Noels \& Sauval 1996)
was adopted in all cases as a solar  reference value.  
The error bars shown in the figure were derived by error propagation of  the
Zn\,II and H\,I column density errors quoted by the authors and of the meteoritic 
abundance error. 

No systematic differences appear to be present between the ZnII column densities 
obtained  from 4-m class telescopes (e.g.  Pettini et al. 1994, 1997,
1999) and those measured with the Keck
telescope (e.g.   Lu et al. 1996, Prochaska \& Wolfe 1996, 1997, 1999). 
In the few cases when   a re-determination of the same Zn II column density
with higher resolution and signal-to-noise ratio is available the difference   is well
within the observational errors quoted by the authors. 
The reason for this is probably the weak degree of saturation of the
Zn\,II   doublet which allows accurate column densities to be determined
even with spectra of modest resolution and signal-to-noise ratio.
We believe therefore that the data sample shown in Fig. \ref{Fig1} is
sufficiently homogeneous to perform a statistical analysis. 

Since the dependence of [Zn/H] as a function of $z$ is not known a priori,
we first performed a non parametric correlation analysis
by means of Kendall's $\tau$ using routine {\tt kendl1} of
Press et al. (1992).
We obtain evidence for rank reversal ($\tau = -0.29$), --- i.e. evidence for anti-correlation ---
at 97.2\% confidence level. 
We then performed
a least square linear regression through the \{$z_{\rm abs}$, [Zn/H]\} data points  
and obtained
an anti-correlation with slope $m = -0.33 \pm 0.12 $
and intercept $q = -0.43 \pm 0.24$.  
Student's t tests
indicate that the slope  
and the intercept differ from zero at 98.7\%  
and 92.0\% confidence level, respectively.
The  linear regression results are confirmed by  
a bootstrap analysis with 10,000 samples, which yields
a mean slope $<m> = -0.32 \pm 0.13$ and a mean intercept $<q> = -0.44 \pm 0.25$.

While the evidence for anti-correlation is quite robust,  
the data   show  a   scatter larger than the typical   measurement errors.
The scatter is still above the individual errors
even considering only the Keck sub-sample. Therefore
the dispersion  is genuine and not  due to inhomogeneity of the data. 
An intrinsic scatter is expected
if DLA systems are associated with galaxies of different types and different
chemical evolution histories.  
An additional source of scatter is the random   galactocentric
distances of the gas intercepted, in the presence of metallicity gradients. 

The influence of intrinsic scatter is
a matter of concern in performing the ordinary least squares
regression on the data. Therefore we re-analysed the data sample   with
the BCES method of Akritas \& Bershady (1996) which is ideally suited
to cope with this situation. We find  
a slope   $m_{\rm BCES} = -0.27 \pm 0.11$ and an intercept
 $q_{\rm BCES} = -0.54 \pm 0.20$.  
The mean   slope and intercept of 10,000 bootstrap samples  
analysed with the BCES method are
 $<m_{\rm BCES}>=-0.32 \pm 0.13$ and  $<q_{\rm BCES}> = -0.45 \pm 0.25$. 
The BCES results confirm, within the errors,
the parameters derived from  the ordinary linear regression analysis. 

Albeit small, zinc shows some dust depletion which may 
affect the above analysis. In the Galactic ISM, the typical zinc depletion is
$\simeq$ $-0.19$ dex (Roth \& Blades 1995), while
in  DLA systems is  $\simeq -0.12$ dex owing to the modest
dust-to-gas ratios typical of these absorbers (Vladilo 1998).  
The intercept of the linear regressions
that we find may be therefore underestimated by
$\lsim$ 0.2 dex (probably $\simeq$ 0.1 dex), a systematic error
which is still  within the statistical errors. 
As far as the slope is concerned, we may expect an increase of its
absolute value (i.e. a steepening of the anti-correlation) as a consequence
of correcting for dust depletion.  The reason for this is that
dust-to-gas ratio and metallicity are correlated
in DLA systems (Vladilo 1998) and we may expect a negligible   correction
at high redshift, where the metallicity is low,
but a non negligible correction at low redshift, where the metallicity rises.
In the most extreme case  (i.e. when we consider the
depletion completely negligible at $z=3.5$ and we apply a  
ISM correction of $+0.19$ dex
at $z=0.5$) we estimate that  
$|m|$ would increase by about 0.06 dex.  The real systematic error is 
probably smaller and hence well within the  
statistical errors of the slope.

\section{The column-density weighted metallicity}

The analysis performed in the previous section is appropriate to study the evolution 
of metals in  individual systems. However, 
in order to  estimate the mean cosmic metallicity of the neutral gas
in the universe, the abundance of each DLA system has to be weighted 
by its HI column density (Pei \& Fall 1995, Pettini et al. 1997). 
It is easy to show that this is equivalent to compute the expression
\begin{equation}
\label{meanZ}
<Z> = \log { \sum_i N({\rm Zn\,II})_i \over \sum_i N({\rm H\,I})_i } - 
\log ({\rm Zn/H})_{\odot}  ~,
\end{equation}
where the sums are extended to the DLA systems  in a given redshift bin.
The mean cosmic metallicity 
$<Z>$  is useful for probing models of cosmic chemical evolution
(Pei \& Fall 1995), but it is particularly affected by low-number statistics.
In fact, not only the number of systems available in each redshift bin is limited
but, in addition,  
only the few of them with the highest $N$(HI) give a significant
contribution to $<Z>$. 

We  estimated  $<Z>$ in 4  redshift bins. 
Rather than adopting bins of constant redshift width, we   
binned 4 groups of  DLA systems with adjacent redshifts, each group 
with comparable value of $\sum_i N({\rm H\,I})_i$ and hence comparable
statistical significance. 
We   then determined the effective redshift of each bin 
with the expression
$< z_{\rm abs} > =  \sum_i  \left[ z_{{\rm abs},i}  \, N({\rm H\,I})_i \right]
/ \sum_i  N({\rm H\,I})_i $. 
The resulting 4 data points are shown as  filled diamonds in Fig. \ref{Fig2}. 
The error bars  of  $<Z>$ are estimated  
 according to  Eq. (3) by Pettini et al. (1997). 
A quick look at the figure does not show  evidence for  metallicity evolution.
This result is confirmed by
a regression analysis of the \{$<z_{\rm abs}>,<Z>$\} data points, which yields
$m=-0.13	\pm 0.07$. This slope is lower than the one derived from the
analysis of individual DLA systems and differs from zero only at
79\% confidence level. 

\section{Summary and conclusions}

We find evidence for an anti-correlation between the absolute zinc abundance
[Zn/H]  and the absorption redshift $z_{\rm abs}$
of DLA systems, with a slope $\simeq -0.3 \pm 0.1$  in the   range
$0.5 \lsim z_{\rm abs} \lsim 3.5$. 
The zinc metallicity increases from 
$\approx 3\%$ up to $\approx 25\%$ of the solar value
from  $z_{\rm abs} \simeq 3.5$ to  $z_{\rm abs} \simeq 0.5$.
Should DLA absorbers continue the same trend   also from
$z_{\rm abs} \simeq 0.5$ to 
$z_{\rm abs} \simeq 0$, the typical present-day   metallicity would
be $\approx 35\%$  solar, even though a value as high as $60\%$ is still
within the errors of the intercept.   
Correcting for dust depletion effects would
slightly steepen the anti-correlation, but well within the statistical error of the slope;
the  characteristic present-day  metallicity would rise up to  $\approx 50\%$ solar,
with values as high as $\approx 100\%$ solar still within the errors.   

The slope of the metallicity redshift relation that we derive
is in good agreement with the value recently derived  
by Savaglio et al. (2000).  The literature data base considered
by these authors is larger than ours and  includes different elements in addition to zinc.
However, most of these elements are known to be severely depleted into dust
in the ISM and the results by Savaglio et al. are based on an algorithm that
corrects the  abundances for depletion effects. 
The present results do not require a modeling of   the
elemental depletion patterns nor
assumptions on the intrinsic   abundance patterns
in DLA systems. 

In spite of the correlation with redshift, the zinc metallicities show evidence for
intrinsic scatter. 
Models of galactic chemical evolution have already been able to explain
such scatter by considering the surface brightness and the
formation redshift of the galaxies, as well as the galactocentric distance 
of the gas intercepted 
(Jim\'enez, Bowen \& Matteucci 1999). 
While such analysis  has shown a general consistency between the zinc
observations and the predicted evolutionary tracks, the present results demonstrate
for the first time the evolution of zinc metallicity on pure observational grounds.

\begin{figure}
\psfig{figure=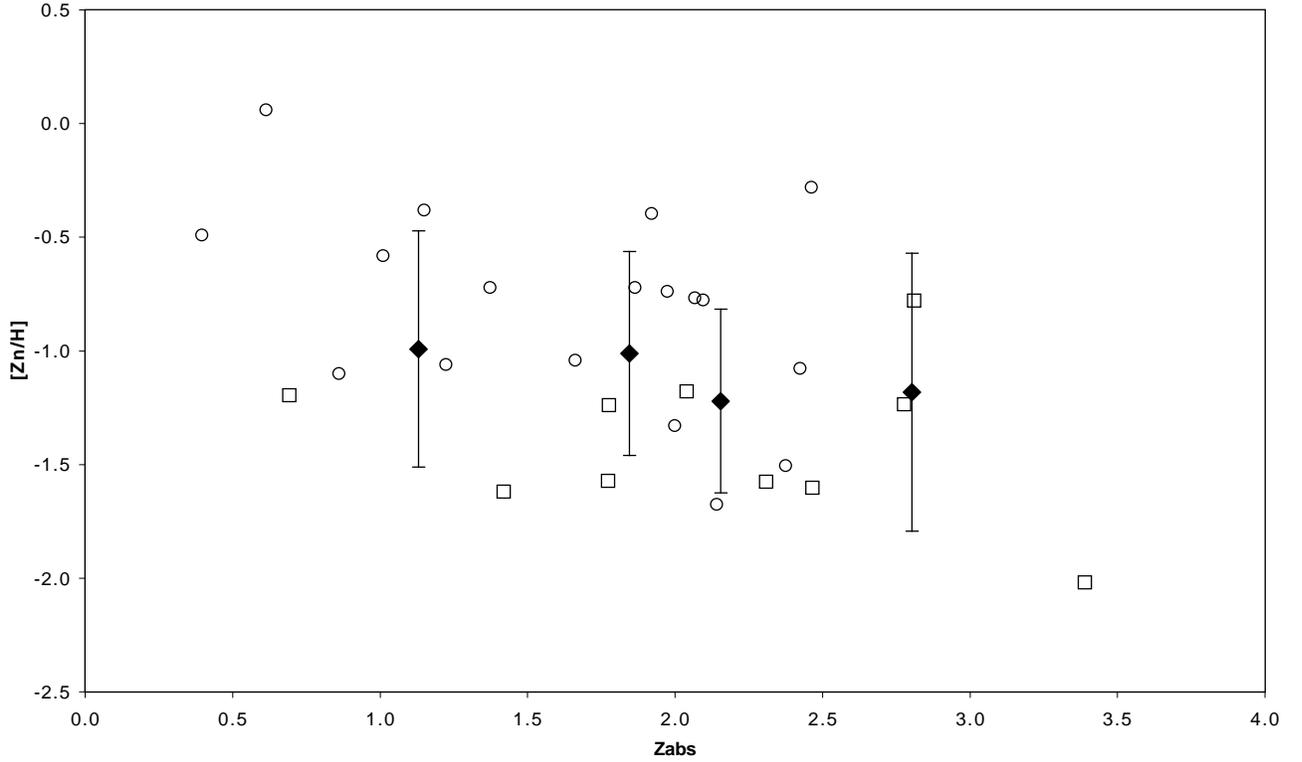,width=17.1cm}
\caption{ 
Filled diamonds: mean weighted metallicity  $<Z>$ 
estimated with Eq.(\ref{meanZ}) for 4 groups of DLA systems 
with comparable value of  $\sum_i N({\rm H\,I})_i$.
Empty symbols:   data points of Fig. 1 labeled according to their
HI column density. Squares and   circles indicate
DLA systems with $N$(HI) $> 10^{21}$ cm$^{-2}$
and $N$(HI) $\leq 10^{21}$ cm$^{-2}$, respectively. }
\label{Fig2}
\end{figure}

While we find evolution of the zinc metallicity of individual systems, we do not find
evolution of the column-density weighted metallicity $<Z>$. 
This is consistent with the results of 
previous studies  by Pettini et al. (1997, 1999). 
The lack of evolution of the zinc mean cosmic metallicity
might be due, at least in part, to the lack of a sufficiently large data base,
since the measurement of $<Z>$ is extremely sensitive to low number
statistics.  
However, there are also reasons to believe that $<Z>$ is affected by 
some selection bias. 
In Fig. \ref{Fig2}  we use   different symbols  for the  systems with
$N$(HI) $> 10^{21}$   cm$^{-2}$ (empty squares) and  
    $N$(HI) $\leq 10^{21}$   cm$^{-2}$   (empty circles). 
One can see from the figure that the systems with  high column density
have, in general, low metallicity ---  an effect originally pointed out by Boiss\'e et al. (1998).
High column density systems are the main contributors
to $<Z>$ and the lack of   such absorbers  with   [Zn/H] $> -1$
at low $z$ tends to hide the global rise
of metallicity with cosmic time.
The lack of  DLA systems of high column density and metallicity at low redshift
is somewhat surprising because (i)
clouds with  $N$(HI) $> 10^{21}$ atoms cm$^{-2}$ and
high metallicity do exist in the disk of our Galaxy and  in
low redshift spirals; (ii)  study of the HI content of the local
universe suggest that spirals should
be the main contributors to the DLA population  
at $z\approx 0$  (Rao \& Briggs 1993).  
Nevertheless, spirals are 
a small fraction of the intervening DLA galaxies observed
in low-$z$ imaging studies (Le Brun et al. 1997; Rao \& Turnshek 1998;
see also refs. in Table 1 by Vladilo 1999).
This deficiency of spirals  suggests the presence of some selection effect. 
Selection effects that can bias the observed population of DLA
absorbers include  QSO obscuration by DLA dust
(Fall \& Pei  1993) and gravitational lensing
 (Smette, Claeskens \& Surdej 1997). However, also the  
surface brightness   of the intervening galaxies and the  
galactocentric distances of the clouds intercepted  
can play a role in affecting  the  observed population. 
As discussed in Vladilo (1999),  these effects generally conspire to 
decrease the fraction of chemical enriched regions in the sample population,
dust obscuration  alone yielding a QSO visual extinction of
$\approx 1$ magnitude  
when  $N$(HI) $> 10^{20.7}$  cm$^{-2}$ at solar metallicity.   
Considering the likely presence of this bias and the severe dependence of $<Z>$
on low number statistics,  
the  lack of   evolution of   $<Z>$ should not be
used to conclude that the mean cosmic metallicity of DLA absorbers does not evolve.  
Comparison between empirical $<Z>$ determinations and model predictions of 
global enrichment of the universe should await a
better understanding of the  role played by  any   selection bias and
a  significant enlargement of the  observational data base.

 \newpage

\end{document}